\documentclass[aps,pra,noshowpacs,twocolumn,reprint,amsmath,amssymb,superscriptaddress,floatfix]{revtex4-1}

\usepackage{color}
\usepackage{amsmath}
\usepackage{graphicx}
\usepackage{dcolumn}
\usepackage{bm}
\usepackage{booktabs}
\usepackage{multirow}

\makeatletter
\newcommand{\rmnum}[1]{\romannumeral #1}
\newcommand{\Rmnum}[1]{\expandafter\@slowromancap\romannumeral #1@}
\makeatother

\begin{document}
\preprint{APS/123-QED}

\title{Observing Movement of Dirac Cones from Single-Photon Dynamics}

\author{Yong-Heng Lu}
\affiliation{Center for Integrated Quantum Information Technologies (IQIT), School of Physics and Astronomy and State Key Laboratory of Advanced Optical Communication Systems and Networks, Shanghai Jiao Tong University, Shanghai 200240, China}
\affiliation{CAS Center for Excellence and Synergetic Innovation Center in Quantum Information and Quantum Physics, University of Science and Technology of China, Hefei, Anhui 230026, China}

\author{Yao Wang}
\affiliation{Center for Integrated Quantum Information Technologies (IQIT), School of Physics and Astronomy and State Key Laboratory of Advanced Optical Communication Systems and Networks, Shanghai Jiao Tong University, Shanghai 200240, China}
\affiliation{CAS Center for Excellence and Synergetic Innovation Center in Quantum Information and Quantum Physics, University of Science and Technology of China, Hefei, Anhui 230026, China}

\author{Yi-Jun Chang}
\affiliation{Center for Integrated Quantum Information Technologies (IQIT), School of Physics and Astronomy and State Key Laboratory of Advanced Optical Communication Systems and Networks, Shanghai Jiao Tong University, Shanghai 200240, China}
\affiliation{CAS Center for Excellence and Synergetic Innovation Center in Quantum Information and Quantum Physics, University of Science and Technology of China, Hefei, Anhui 230026, China}

\author{Zhan-Ming Li}
\affiliation{Center for Integrated Quantum Information Technologies (IQIT), School of Physics and Astronomy and State Key Laboratory of Advanced Optical Communication Systems and Networks, Shanghai Jiao Tong University, Shanghai 200240, China}
\affiliation{CAS Center for Excellence and Synergetic Innovation Center in Quantum Information and Quantum Physics, University of Science and Technology of China, Hefei, Anhui 230026, China}

\author{Wen-Hao Cui}
\affiliation{Center for Integrated Quantum Information Technologies (IQIT), School of Physics and Astronomy and State Key Laboratory of Advanced Optical Communication Systems and Networks, Shanghai Jiao Tong University, Shanghai 200240, China}
\affiliation{CAS Center for Excellence and Synergetic Innovation Center in Quantum Information and Quantum Physics, University of Science and Technology of China, Hefei, Anhui 230026, China}

\author{Jun Gao}
\affiliation{Center for Integrated Quantum Information Technologies (IQIT), School of Physics and Astronomy and State Key Laboratory of Advanced Optical Communication Systems and Networks, Shanghai Jiao Tong University, Shanghai 200240, China}
\affiliation{CAS Center for Excellence and Synergetic Innovation Center in Quantum Information and Quantum Physics, University of Science and Technology of China, Hefei, Anhui 230026, China}

\author{Wen-Hao Zhou}
\affiliation{Center for Integrated Quantum Information Technologies (IQIT), School of Physics and Astronomy and State Key Laboratory of Advanced Optical Communication Systems and Networks, Shanghai Jiao Tong University, Shanghai 200240, China}
\affiliation{CAS Center for Excellence and Synergetic Innovation Center in Quantum Information and Quantum Physics, University of Science and Technology of China, Hefei, Anhui 230026, China}

\author{Hang Zheng}
\thanks{hzheng@sjtu.edu.cn}
\affiliation{Laboratory of Artificial Structures and Quantum Control (Ministry of Education), School of Physics and Astronomy, Shanghai Jiao Tong University, Shanghai 200240, China}

\author{Xian-Min Jin}
\thanks{xianmin.jin@sjtu.edu.cn}
\affiliation{Center for Integrated Quantum Information Technologies (IQIT), School of Physics and Astronomy and State Key Laboratory of Advanced Optical Communication Systems and Networks, Shanghai Jiao Tong University, Shanghai 200240, China}
\affiliation{CAS Center for Excellence and Synergetic Innovation Center in Quantum Information and Quantum Physics, University of Science and Technology of China, Hefei, Anhui 230026, China}	
	
\date{\today}
	
\maketitle

\textbf{Graphene with honeycomb structure, being critically important in understanding physics of matter, exhibits exceptionally unusual half-integer quantum Hall effect and unconventional electronic spectrum with quantum relativistic phenomena. Particularly, graphene-like structure can be used for realizing topological insulator which inspires an intrinsic topological protection mechanism with strong immunity for maintaining coherence of quantum information. These various peculiar physics arise from the unique properties of Dirac cones which show high hole degeneracy, massless charge carriers and linear intersection of bands. Experimental observation of Dirac cones conventionally focuses on the energy-momentum space with bulk measurement. Recently, the wave function and band structure have been mapped into the real-space in photonic system, and made flexible control possible. Here, we demonstrate a direct observation of the movement of Dirac cones from single-photon dynamics in photonic graphene under different biaxial strains. Sharing the same spirit of wave-particle nature in quantum mechanics, we identify the movement of Dirac cones by dynamically detecting the edge modes and extracting the diffusing distance of the packets with accumulation and statistics on individual single-particle registrations. Our results of observing movement of Dirac cones from single-photon dynamics, together with the method of direct observation in real space by mapping the band structure defined in momentum space, pave the way to understand a variety of artificial structures in quantum regime.}\\
		
Dirac cones lead to fascinating properties of graphene structure which exhibit quantum relativistic dynamics and the robust unidirectional edge states in topological insulators \cite{gra_1,gra_2,gra_3}. Around Dirac cones, conduction and valence bands intersect linearly at a degenerate point resulting in the linear energy band and massless relativistic transport of electrons as quasiparticles \cite{gra_4}. A pair of degenerate points, behaving as topological defects in the band structure with associated Berry phase of $\pi$ and $-\pi$, guarantees the stability and robustness against small perturbations and disorders, which inspires topologically protected quantum computing \cite{computing}. The investigations of Dirac cones and graphene have pushed the boundaries of our understanding on emerging physics and materials, for both fundamental and device levels \cite{app_1,app_2,app_3,app_4}.

Recently, varieties of artificial systems have enabled the imitation and exploration of graphene and Dirac cones using the honeycomb structure with controllable parameters and observables \cite{model1,model2,model3,model4,model5,model6,model7,model8,model9,model10,model11,model12,model13,model14,model15,model16} beyond traditional graphene itself where those are not easily accessible. For instance, observation of Dirac cones has been realized in cold atom gases \cite{model13}, acoustic waves system \cite{model14}, polaritons system \cite{model15} and photonic resonators system \cite{model16}. The band structure related to Dirac cones has been widely investigated in the energy-momentum space using the bulk measurement with measured density and dispersion pattern \cite{model1,model13,model14,model15,model16}. The manipulation and manifestation of the wave function and band modes can be appropriately mapped to the robustly certain observables in 1D real space \cite{Zeuner2015,wang2019}. The mapping approach can be extended to higher dimension for dynamical detection of high-complex artificial structures, which, fascinatingly, can make it possible to directly observe the movement of Dirac cones in real space.

As a typical physical wave, light is well compatible to graphene theory, meanwhile as a stream of photons, light can be manipulated and detected down to single-particle level. Such a unique feature of light associated with photonic graphene theory suggests an emerging crossover of material and quantum physics, enabling investigation of materials in quantum regime.  

Here, we map the band structure of photonic graphene from momentum space to real space and successfully observe the movement of Dirac cones from single-photon dynamics. We construct the isotropic graphene and the two anisotropic cases under different biaxial strains. We identify the movement of Dirac cones in all cases via dynamically detecting the edge modes and extracting the diffusing distance of the packets. 
	
		\begin{figure}[]
			\centering
			\includegraphics[width=1\columnwidth]{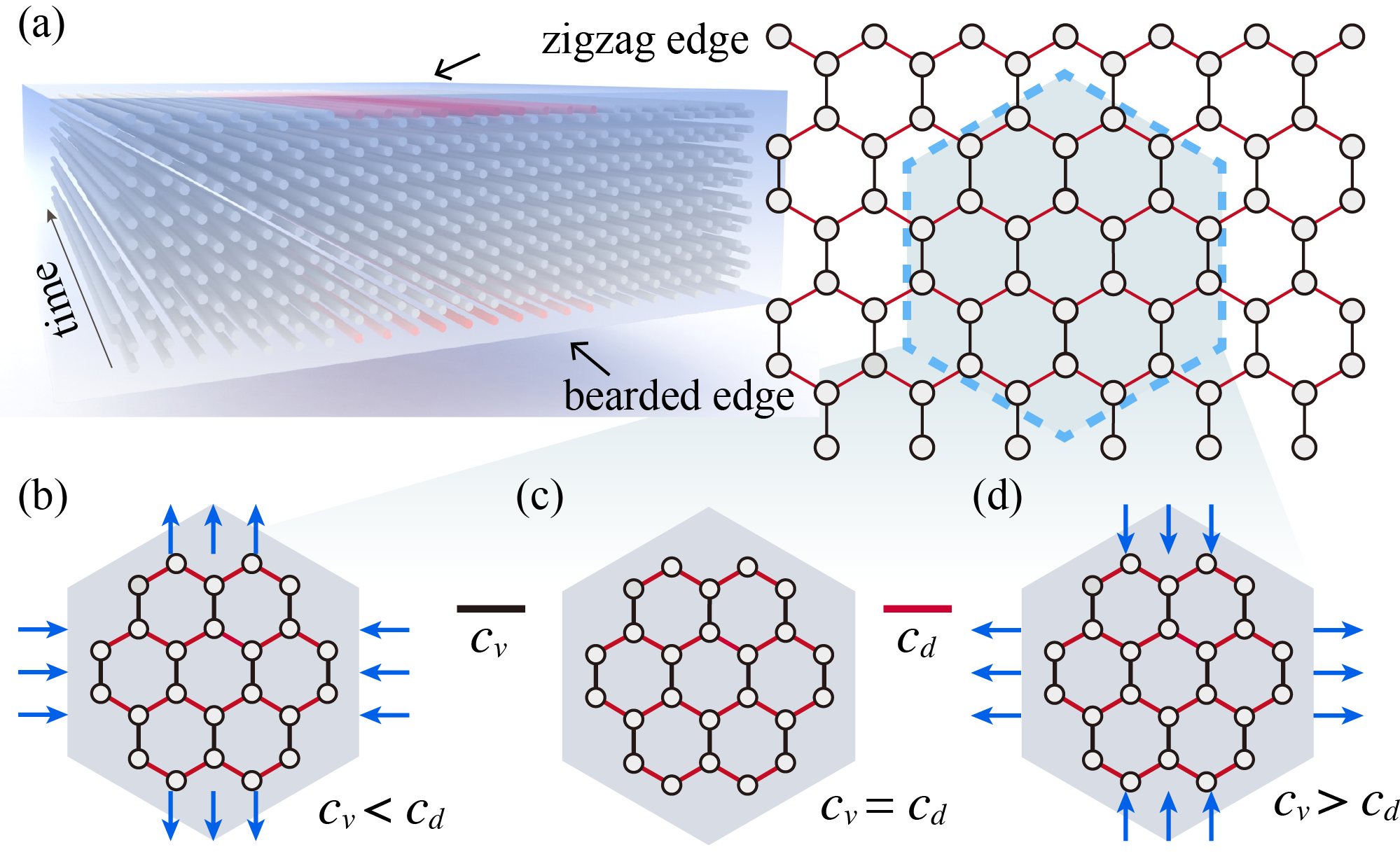}\\
			\caption{\textbf{Construction and modeling of photonic graphene with biaxial strains}  (\textbf{a}) Schematic diagram of large-scale integrated photonic graphene on silica chip. Single photons are injected into the entry waveguide in the bottom boundary (bearded edge) and the top boundary (zigzag edge). (\textbf{b}) Anisotropic graphene model under the compressive strain in the horizontal direction and tensile strain in the vertical direction, resulting in $c_{v}<c_{d}$. (\textbf{c}) Isotropic graphene model with $c_{v}=c_{d}$. (\textbf{d}) Anisotropic graphene model under the tensile strain in the horizontal direction and compressive strain in the vertical direction, resulting in $c_{v}>c_{d}$.}
			\label{f1}
		\end{figure}
	

\begin{figure}
	\centering
	\includegraphics[width=1\columnwidth]{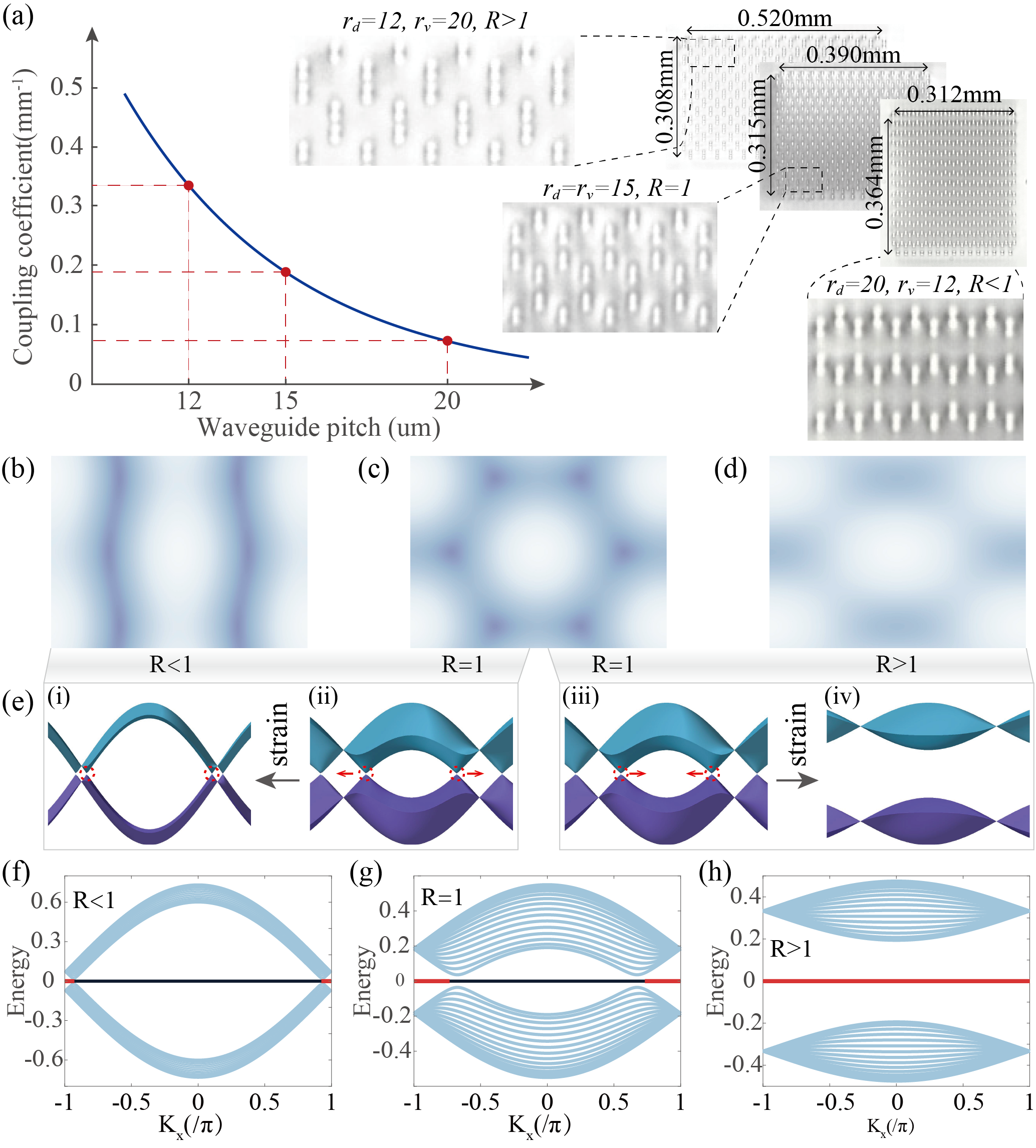}\\
	\caption{\textbf{Moving of Dirac cones via introduction of the anisotropy  
		} (\textbf{a}) Microscopic image of cross section of three types of photonic graphene with the construction parameters. Parameters $d_{d}$ and $d_{v}$ are marked by the red dots in the relationship between the separation of adjacent waveguides and the nearest coupling coefficient. (\textbf{b})-(\textbf{d}) show the energy dispersion with Dirac cones or without Dirac cones in three types cases. (\textbf{e}) Band structure of the three types cases along the $k_{x}$ parametric plane. Red arrows indicate the movement of the Dirac cones when the anisotropy is introduced. (\textbf{f}) Edge band structure of three types cases in limit. Bulk modes, bearded mode and zigzag mode are denoted by wathet blue, dark blue and red, respectively. }
	\label{f2}
\end{figure}

The photonic graphene with the honeycomb structure\cite{model3,model5} holds the zigzag termination and bearded termination on the top and bottom boundary, arranged as depicted in Fig.1a. Applying the `tight-binding approximation' to the `photonic graphene' \cite{tb}, the Hamiltonian can be described by
	\begin{equation}
	\label{eq1}
		H^{TB}=\sum_{i\ne j}^{N} {c_{i,j}a_{i}^{\dagger}a_{j}+H.c.}
	\end{equation}	
where $N$ is the total number of the waveguides and $c_{i,j}$ denotes the coupling constant between the nearest-neighbor waveguides $i$ and $j$. Furthermore, we note that the coupling constant $c_{i,j}$ concludes two parts: vertical coupling $c_{v}$ and diagonal coupling $c_{d}$, as depicted in Fig.1b-d. 

In the isotropic graphene, $c_{v}=c_{d}$, the lattice is hexagonal and the tunnelling is equivalent in all directions, which is preserved by the six-fold symmetry \cite{gra_1}. In contrast, introducing the anisotropy along the vertical direction and diagonal direction into the graphene leads to the broken six-fold symmetry and level crossing with the deformation of the bandstructure \cite{ani1,ani2,ani3}. Here, the anisotropy can be introduced by applying the biaxial strains to the graphene, as shown in Fig.1b-d. The horizontal direction is implemented with the tensile (compressive) strain while the vertical direction is implemented with the compressive (tensile) strain, which is corresponding to the case of $c_{v}>c_{d}$ ($c_{v}<c_{d}$), respectively. Because the diffraction of light in photonic graphene can be described by the paraxial discrete Schr$\ddot{o}$dinger equation, the propagation of photons through the three types of graphene are all governed by:                                                                                                                                                                                                                                                  
\begin{equation}
\label{eq2}
i\partial_{z} \psi_{i}(z)=H^{TB}\psi_{j}(z)
\end{equation}

\begin{figure*}
	\centering
	\includegraphics[width=2\columnwidth]{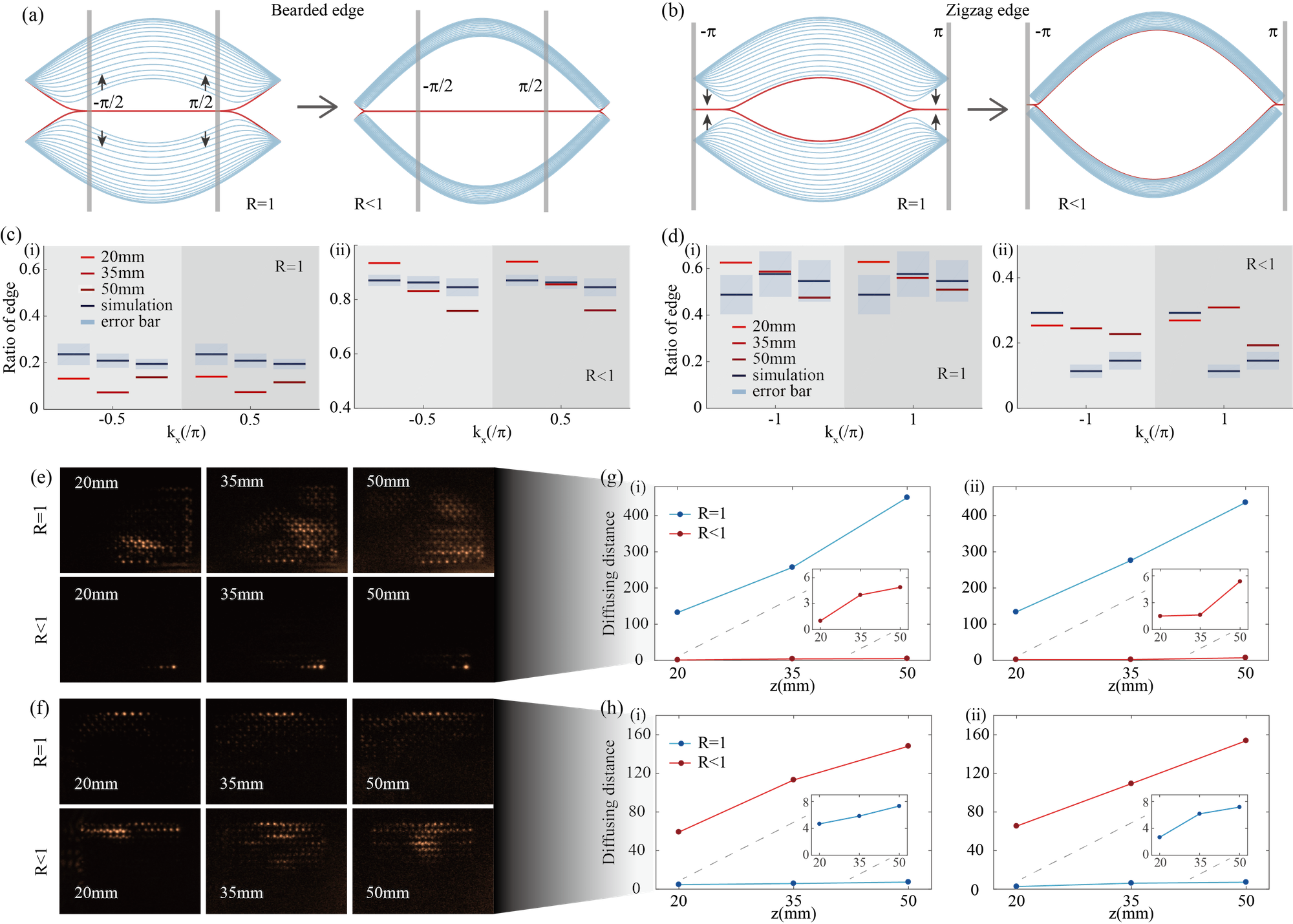}\\
	\caption{\textbf{Experimental results for observing Dirac cones for $R<1$} (\textbf{a}-\textbf{b}) Bearded edge band structure and zigzag edge band structure. The selected region of Bloch wavevectors is marked by the grey belt. The edge modes are marked by red lines. (\textbf{c}-\textbf{d}) Ratio of the photons staying in the edge for $k_{x}=-\frac{\pi}{2}$, $k_{x}=\frac{\pi}{2}$ and $k_{x}=-\pi$, $k_{x}=\pi$ for exciting the bearded edge and zigzag edge. (\textbf{e}) Dynamical distribution patterns of the output facet for exciting bearded edge at $k_{x}=\frac{\pi}{2}$. (\textbf{f}) Dynamical distribution patterns of the output facet for exciting zigzag edge at $k_{x}=\pi$. (\textbf{g}) Dynamical diffusing distance for $k_{x}=-\frac{\pi}{2}$(\rmnum{1}) and $k_{x}=\frac{\pi}{2}$(\rmnum{2}). (\textbf{h}) Dynamical diffusing distance for $k_{x}=-\pi$(\rmnum{1}) and $k_{x}=\pi$(\rmnum{2}).}
	\label{f3}
\end{figure*}

The three types of photonic graphene, individually containing 449 waveguides, are integrated in the silica chip by using the three dimensional femtosecond laser direct writing technique which allows freely written lattices with various geometry \cite{w1,w2,qw_tang,tang_2018}. In the waveguide system, according to the nonlinear relationship between next-nearest coupling strength and nearest-neighbour distance, the case for $c_{v}>c_{d}, c_{v}=c_{d}$ and $c_{v}<c_{d}$ can be denoted by $R<1, R=1$ and $R>1$ as shown in Fig.2(a) (see Methods). 

 As shown in Fig.2b-h, the band structure can be modulated by introducing the anisotropy. In case for $R=1$, it exhibits double-conical Dirac band structure with the Dirac points. In contrast, in case for $R<1$, a pair of Dirac cones moves apart from each other along the axis of $k_{x}$ and approaches the boundary of the Brillouin zone, as shown in Fig.2e(\rmnum{1})-(\rmnum{2}). Fig.2f shows that the applied anisotropy further broadens the region of bearded mode and narrows the region of zigzag mode comparing to that in isotropic case. Therefore, one the one hand, the moving of the Dirac cones in the case of $R<1$ leads to the larger and wider band gap between the bulk and bearded edge mode with the smaller band gap between the bulk and zigzag edge mode in edge band structures. On the other hand, for $R>1$, a pair of Dirac cones move towards each other and show the possess of moving, merging and annihilating related to the opening band gap and phase transition, resulting in zigzag edge mode occupying all of the Brillouin zone with the disappearing of bearded edge mode. One should note that the opening band gap in graphene requires large strain, even exceeding the order of $20\%$ \cite{ani1,model15}, which is attributed to the deformations and anisotropy by applying strain instead of the inner disorders or fluctuation of the graphene. Thus, it means that probing the band structure can be a hint for reflecting the moving of the Dirac cones in the anisotropic case. 

\begin{figure*}
	\centering
	\includegraphics[width=1.95\columnwidth]{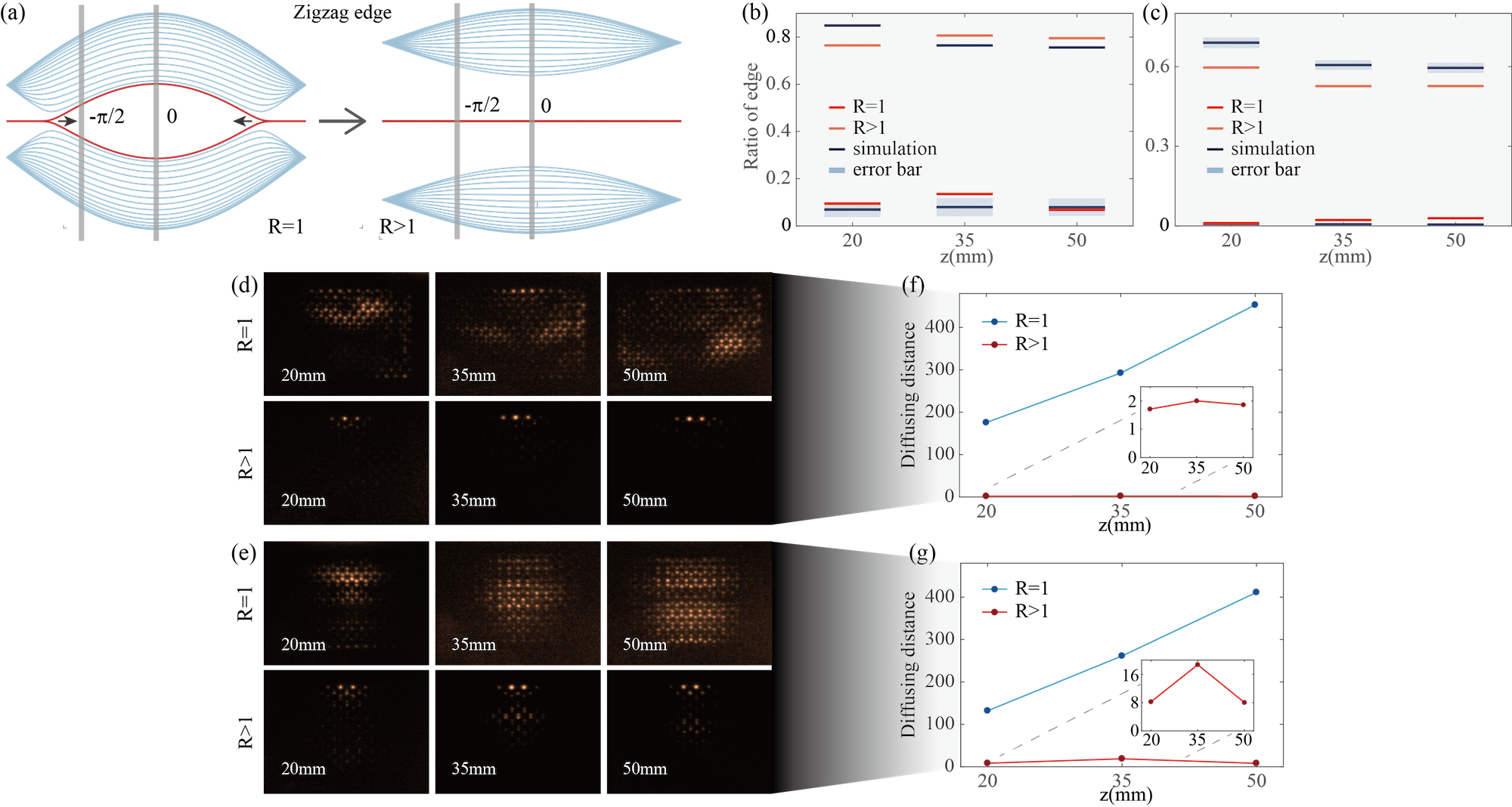}\\
	\caption{\textbf{Experimental results for observing Dirac cones for $R>1$} (\textbf{a}) Zigzag edge band structure. The selected region of the Bloch wavevectors is marked by the grey belt. The edge modes are marked by red lines. Ratio of the photons staying in the edge at $k_{x}=-\frac{\pi}{2}$ (\textbf{b}) and $k_{x}=0$ (\textbf{c}). Dynamical distribution patterns of the output facet for zigzag modes at $k_{x}=-\frac{\pi}{2}$ (\textbf{d}) and $k_{x}=0$ (\textbf{e}). (\textbf{f}) Dynamical diffusing distance for $k_{x}=-\frac{\pi}{2}$. (\textbf{g}) Dynamical diffusing distance for $k_{x}=0$.}
	\label{f4}
\end{figure*}
	
In experiment, we prototype the 2D photonic lattices with different propagation distances varying from 20 to 50 mm with step size of 15 mm. Here, the z-axis replaces the temporal coordinate. We prepare heralded single photons by conditionally triggering one photon from a photon pair via spontaneous parametric down conversion. The probability distribution of single photons can be dynamically imaged by an Intensified Charge-Coupled Device (ICCD) camera at the output facet of the lattices (see Methods). 

For mapping the band structure influenced by the Dirac cones into the observables in real space, it is convenient to excite bearded and zigzag edge modes at different Bloch wavevectors, which corresponds to the manifestation of wave function. As shown in the Fig.3a-b, for the case of $R<1$, Bloch wavevector $\vert \frac{\pi}{2}\vert$ for bearded edge and $\vert\pi\vert$ for zigzag edge are both sensitive to the deformation of band structure with enlarging or narrowing the width of band gap which reflect the coupling or decoupling processes between the edge modes and bulk modes. Therefore, exciting bearded and zigzag modes can lead to very distinct phenomena in real space simultaneously under the same deformation. 

For anisotropic case $R<1$, we inject the heralded single photons with $\vert \frac{\pi}{2}\vert$ Bloch wavevector into the bottom entrances of the lattices to excite the bearded mode. Fig.3c shows that the ratio of photons confining in the edge is steadily low in $R=1$ and distinctly keeps high without being affected by the increasing evolution distance in the case of $R<1$. Meanwhile, we also excite the zigzag modes with $\vert\pi\vert$ Bloch wavevector. Different from the performance of bearded modes, there are strong confinement in case $R=1$ but weak confinement in case $R<1$ in Fig.3d. Thus, simultaneous observation of the creation of bearded edge modes, together with destruction of zigzag edge modes in case $R<1$, well exhibits the movement of Dirac cones with the deformed band structure.

Similar to the transport properties of electrons in graphene, the evolution of photons as the carriers in photonic graphene deserves to be explored. After dynamically probing the output patterns which are shown in Fig.3e-f, we evaluate and compare the evolution of photons in cases of $R<1$ and $R=1$ by extracting the center of packets and calculating the diffusing distance (see Methods). Fig.3g shows high and increasing the diffusing distance for the case of $R=1$ and nearly zero diffusing distance for the case of $R<1$ while exciting the bearded edge at $\vert \frac{\pi}{2}\vert$ wavevector. In contrast, the photons prefer to be confined in the edge for the case of $R=1$ and diffract into the bulk with increasing evolution distances for the case of $R<1$ while exciting the zigzag edge at $\vert\pi\vert$ wavevector, as shown in Fig.3h, which is consistent with the deformation of band structure induced by the moving of Dirac cones.

In order to examine the reliability and universality of our identification and observation, we further experimentally investigate the case for $R>1$. As shown in Fig.4a, there is the opening band gap for zigzag mode with merged Dirac cones. The zigzag modes can be coupled into the bulk for the case of $R=1$ and decoupled from the bulk for the case of $R>1$ at $-\frac{\pi}{2}$ and $0$. We inject the heralded single photons with two different Bloch wavevectors $-\frac{\pi}{2}$ and $0$ into the top entrances of the lattices to excite zigzag mode. The ratio of edge for the case of $R>1$ at different evolution distances is continuously much higher than that of the case of $R=1$ (see Fig4.b-c). Moreover, it is also very apparent that the dynamical evolution of the photons can be distinctly distinguished between the two cases (see Fig4.d-g). 
	
In summary, we present direct observation of the movement of Dirac cones from single-photon dynamics in photonic graphene fabricated with femtosecond laser direct writing technique. After constructing the isotropic and two anisotropic graphene under the different biaxial strain, we identify the movement of Dirac cones in all cases via detecting the ratio of photons in the edge and extracting diffusing distance along the evolution of photons. The interplay between Dirac cones, edge modes and single-photon evolution provides a new route to simulate and unveil complex condensed matter physics in real space and in genuine quantum regime.

With the successful experimental observation and validation here, it is of great interest to generalize our approach of dynamical observation to the other attractive models, including Lieb-lattices \cite{lieb1,lieb2,lieb3,lieb4} with localized states in the continuum, Kagome-breathing-lattices \cite{ka1}, fractal topological photonic quasicrystals \cite{fraction}, high-dimensional system with disorder and defects \cite{defect1}. Along with manipulation of Dirac cones, edge modes coupling or decoupling from bulk modes, band gap enlarging and evolution of photons, our experimental approaches may provide new sight on using photonic systems to engineer the analogous effects for solving complex and hard problems in condensed matter. The elements of single-photon dynamics and their protection in the honeycomb lattices may inspire developing entirely new capacities of engineering quantum states in the different artificial structures, especially topological-protected mechanics \cite{topo1,topo2,topo3}. 

        \subsection*{Acknowledgments}
The authors thank Jian-Wei Pan for helpful discussions. This research was supported by the National Key R\&D Program of China (2019YFA0308700, 2017YFA0303700), the National Natural Science Foundation of China (11690033, 61734005, 11761141014), the Science and Technology Commission of Shanghai Municipality (STCSM) (17JC1400403), and the Shanghai Municipal Education Commission (SMEC) (2017-01-07-00-02- E00049). X.-M.J. acknowledges additional support from a Shanghai talent program.\\
	
	\subsection*{Methods}
{\bf Fabrication of integrated photonic graphenes:}
	The isotropic photonic graphene and two anisotropic graphenes are all well fabricated and integrated in a borosilicate glass substrate (refractive index $n_0=1.514$). Each has 449 waveguides in the lattices in total. The adjacent waveguides interact with each other via evanescent light coupling. Based on the nonlinear relation between coupling coefficients and the separation between two adjacent waveguides, we can freely design the different types of honeycomb structure for isotropic and anisotropic fashion. Spatial light modulator (SLM) are fed by a $513$ $nm$ femtosecond laser (up conversion from $1026$ $nm$ pump laser; $10$ $W$; $290$ $fs$ pulse duration; $1$ $MHz$ repetition rate ) to shape the laser beam with creating burst trains in both time and spatial domain. Then, the pulse is used to focus onto the borosilicate substrate with a 50$X$ objective lens with 0.55 numerical aperture at the constant velocity of $10$ $mm/s$. Due to the large size of lattice, the uniformity of waveguides at different depth need to be improved. Here, uniformity and depth-independence of the large array can be guaranteed by the SLM and power compensation. 
	
In the waveguide system, according to the nonlinear relationship between next-nearest coupling strength and nearest-neighbour distance, the case for $c_{v}>c_{d}, c_{v}=c_{d}$ and $c_{v}<c_{d}$ means $r_{v}<r_{d}, r_{v}=r_{d}$ and $r_{v}>r_{d}$ with different fabricated constructions. Thus, the inequivalence between $r_{v}$ and $r_{d}$ represents the strain and anisotropy. For convenience, we note that $R=r_{v}/r_{d}$. Thus, the case for $r_{v}<r_{d}, r_{v}=r_{d}$ and $r_{v}>r_{d}$ can be denoted by $R<1, R=1$ and $R>1$.\\

{\bf Imaging pattern using heralded single photon:}
We pump a periodically-poled KTP (PPKTP) crystal by a $405$ $nm$ diode laser. The $810$ $nm$ photon pair can be generated via spontaneous parametric down conversion during the pumping processes. One generated photon can be regarded as signal photon and the other one can be regarded as idler one. After generated photon pairs passing the long band-pass filter and a polarized beam splitter, the two components can be purified. Then, we collect the idler photon by the avalanche photo diode (APD) as the trigger, and couple the signal photon into a single-mode optical fibre as heralded single photon source ready to be injected into the photonic graphene. With the heralded single photon source and triggered single-photon imaging, the captured distribution pattern at the output facet of photonic graphene will be able to faithfully reflect the investigated single-photon dynamics. 
	
After the photons launching from single-mode optical fibre, we use cylindrical lens to reshape the probe photons to maintain the shape: narrow along y-axis and wide along x-shape. The narrow-strip-shaped probe photons are able to excite the selected edge (bottom or top edge) of the lattices precisely. The incident angle of probe photons can be controlled in the spectrum plane using a 4-$f$ system (two lens with focal length $f$ at a distance of 2$f$). Then, the photons are focused into the lattices by a 20$X$ objective lens. The translation stages, possessing smooth continuous motion and long-term stability, can precisely control the connection between photons and lattices to reduce the coupling loss. After propagating through the lattices, outgoing photons are shaped by the 16$X$ objective lens and the output is imaged by a triggered ICCD camera. For ICCD camera, we set the time delay of $20$ $ns$ and a gate width of $10$ $ns$ to compensate the time difference experienced by the signal and idler photons. We capture each dynamical evolution pattern by using ICCD camera after accumulating in the external triggering mode for $1000$ $s$.\\
	
{\bf Estimation of diffusing distance:}
The appearance of the edge modes with band gap can be induced by the movement of the Dirac cones in the band structure, which results in localization or diffraction of the photons. Thus, the moving of the Dirac cones can be distinguished and observed by extracting diffusing distance of photons during the evolution. Here, we employ the center recognition algorithm to analyze the obtained evolution patterns. Firstly, we sign the position where photons are injected as the reference and the position is marked by $h_{0}$. Then, we identify the centroid of the cluster of photons and their positions are marked by $h_{1}$. The diffusing distance can be estimated by $\Delta d = h_{1}-h_{0}$. The diffusing distance of photons shows the diffraction into bulk and the localization in edge, which further distinctly reveals and verifies the movement of the Dirac cones.
	


\begin{thebibliography}{99}
		
\bibitem{gra_1} Castro Neto, A. H., Guinea, F., Peres, N. M. R., Novoselov, K. S. \& Geim, A. K. The electronic properties of graphene. \emph{Rev. Mod. Phys.} \textbf{81}, 109–162 (2009).	

\bibitem{gra_2} Das Sarma, S., Adam, S., Hwang, E. H. \& Rossi, E. Electronic transport in
two-dimensional graphene. \emph{Rev. Mod. Phys.} \textbf{83}, 407–470 (2011)

\bibitem{gra_3} Hasan, M. Z. \& Kane, C. L. Topological insulators. \emph{Rev. Mod. Phys.} \textbf{82}, 3045-3067 (2010).

\bibitem{gra_4} Novoselov, K. S. \emph{et al.} Two-dimensional gas of massless Dirac fermions in
 graphene. \emph{Nature} \textbf{438}, 197–200 (2005).
 
\bibitem{computing} Moore, J. E. The birth of topological insulators. \emph{Nature (London)} \textbf{464}, 194 (2010).

\bibitem{app_1} Zhang, Y., Tan, J. W., Stormer, H. L. \& Kim, P. Experimental observation of the
quantum Hall effect and Berry’s phase in graphene. \emph{Nature} \textbf{438}, 201–204 (2005)

\bibitem{app_2} Novoselov, K. S. \emph{et al.} Unconventional quantum Hall effect and Berry’s phase of 2$\pi$ in bilayer graphene. \emph{Nature Phys.} \textbf{2}, 177–180 (2006).

\bibitem{app_3} Butler, S. Z. \emph{et al.} Progress, challenges, and opportunities in two-dimensional materials beyond graphene. \emph{ACS Nano} \textbf{7}, 2898–2926 (2013).

\bibitem{app_4} Ferrari, A. C. \emph{et al.} Science and technology roadmap for graphene, related
two-dimensional crystals, and hybrid systems. \emph{Nanoscale} \textbf{7}, 4598–4810 (2015)

\bibitem{model1} Polini, M., Guinea, F., Lewenstein, M., Manoharan, H. C., \& Pellegrini, V. Artificial honeycomb lattices for electrons, atoms and photons. \emph{Nature Nanotech.} \textbf{8}, 625–633 (2013).

\bibitem{model2} Peleg, O., Bartal, G., Freedman, B., Manela, O., Segev, M., \& Christodoulides, D.N., \emph{Phys. Rev. Lett.} \textbf{98}, 103901 (2007).

\bibitem{model3} Bahat-Treidel, O. \emph{et al.} Klein tunneling in deformed honeycomb lattices.
\emph{Phys. Rev. Lett.} \textbf{104}, 063901 (2010).

\bibitem{model4} Rechtsman, M. C. \emph{et al.} Topological creation and destruction of edge states in photonic graphene. \emph{Phys. Rev. Lett.} \textbf{111}, 103901 (2013).

\bibitem{model5} Rechtsman, M. C. \emph{et al.} Strain-induced pseudomagnetic field and photonic
Landau levels in dielectric structures. \emph{Nature Photon.} \textbf{7}, 153–158 (2013).

\bibitem{model6} Plotnik, Y. \emph{et al.} Observation of unconventional edge states in ‘photonic
graphene’. \emph{Nature Mat.} \textbf{13}, 57–62 (2014).

\bibitem{model7} De Simoni, G., Singha, A., Gibertini, M., Karmakar, B., Polini, M., Piazza, V.,  Pfeiffer, L. N., West, K. W., Beltram, F., \& Pellegrini, V. Delocalized-localized transition in a semiconductor two-dimensional honeycomb lattice. \emph{Appl. Phys. Lett.} \textbf{97}, 132113 (2010).

\bibitem{model8} Gomes, K. K., Mar, W., Ko, W., Guinea, F. \& Manoharan, H. C. Designer Dirac fermions and topological phases in molecular graphene. \emph{Nature} \textbf{483}, 306–310 (2012).

\bibitem{model9} Bellec, M., Kuhl, U., Montambaux, G. \& Mortessagne, F. Topological transition of Dirac points in a microwave experiment. \emph{Phys. Rev. Lett.} \textbf{110}, 033902 (2013).

\bibitem{model10} Bellec, M., Kuhl, U., Montambaux, G. \& Mortessagne, F. Tight-binding couplings in microwave artificial graphene. \emph{Phys. Rev. B} \textbf{88}, 115437 (2013).

\bibitem{model11} Mili$\acute{c}$evi$\acute{c}$, M. et al. Edge states in polariton honeycomb lattices. \emph{2D Mater.} \textbf{2}, 034012 (2015).

\bibitem{model12} Yu, S. \emph{et al.} Surface phononic graphene. \emph{Nat. Mater.} \textbf{15}, 1243–1247 (2016).

\bibitem{model13} Tarruell, L., Greif, D., Uehlinger, T., Jotzu, G., \& Esslinger, T. Creating, moving and merging Dirac points with a Fermi gas in a tunable honeycomb lattice. \emph{Nature (London)} \textbf{483}, 302 (2012).

\bibitem{model14} Torrent, D. \& S$\acute{a}$nchez-Dehesa, J. Acoustic analogue of graphene: observation of Dirac cones in acoustic surface waves. \emph{Phys. Rev. Lett.} \textbf{108}, 174301 (2012).

\bibitem{model15} Jacqmin, T. \emph{et al.} Direct observation of Dirac cones and a fatband in a
honeycomb lattice for polaritons. \emph{Phys. Rev. Lett.} \textbf{112}, 116402 (2014).

\bibitem{model16} Peng, S. \emph{et al.} Probing the band structure of topological silicon photonic lattices in the visible spectrum. \emph{Phys. Rev. Lett.} \textbf{122}, 117401 (2019).

\bibitem{Zeuner2015} Zeuner, J. M. \emph{et al.} Observation of a topological transition in the bulk of a non-Hermitian system. \emph{Phys. Rev. Lett.} \textbf{115}, 040402 (2015).

\bibitem{wang2019} Wang, Y. \emph{et al.} Direct Observation of Topology from Single-Photon Dynamics. \emph{Phys. Rev. Lett.} \textbf{122}, 193903 (2019).


\bibitem{tb} Kohmoto, M. \& Hasegawa, Y. Zero modes and edge states of the honeycomb lattice. \emph{Phys. Rev. B} \textbf{76}, 205402 (2007).


\bibitem{ani1} Pereira, V.M., Castro Neto, A. H. \& Peres, N.M. R. Tight-binding approach to uniaxial strain in graphene. \emph{Phys. Rev. B} \textbf{80}, 045401 (2009).

\bibitem{ani2} Delplace, P., Ullmo, D., \& Montambaux, G. Zak phase and the existence of edge states in graphene. \emph{Phys. Rev. B} \textbf{84}, 195452 (2011).

\bibitem{ani3} Hasegawa, Y. Hasegawa \& Kishigi, K. Merging Dirac points and topological phase transitions in the tight-binding model on the generalized honeycomb lattice. \emph{Phys. Rev. B} \textbf{86}, 165430 (2012).

\bibitem{w1} Davis, K. M., Miura, K., Sugimoto, N. \& Hirao, K. Writing waveguides in glass with a femtosecond laser. \emph{Opt. Lett.} \textbf{21}, 1729-1731 (1996).

\bibitem{w2} Szameit, A., Dreisow, F., Pertsch, T., Nolte, S. \& T$\ddot{u}$nnermann, A. Control of directional evanescent coupling in fs laser written waveguides. \emph{Opt. Express} \textbf{15}, 1579-1587 (2007).

\bibitem{qw_tang} Tang, H. \emph{et al.} Experimental two-dimensional quantum walk on a Photonic chip. \textit{Sci. Adv.} \textbf{4}, eaat3174 (2018).

\bibitem{tang_2018} Tang, H. \emph{et al.} Experimental quantum fast hitting on hexagonal graphs. \emph{Nat. Photon.} \textbf{12}, 754 (2018).


\bibitem{lieb1} Guzmán-Silva, D. et al. Experimental observation of bulk and edge transport in
photonic Lieb lattices. \emph{New J. Phys.} \textbf{16}, 063061 (2014).

\bibitem{lieb2} Mukherjee, S. \emph{et al.} Observation of a localized flat-band state in a photonic lieb lattice. \emph{Phys. Rev. Lett.} \textbf{114}, 245504 (2015).

\bibitem{lieb3} Vicencio, R. A. \emph{et al.} Observation of localized states in Lieb photonic lattices. \emph{Phys. Rev. Lett.} \textbf{114}, 245503 (2015).

\bibitem{lieb4} Slot, M. R. \emph{et al.} Experimental realization and characterization of an electronic
Lieb lattice. \emph{Nat. Phys.} \textbf{13}, 672–676 (2017).

\bibitem{ka1} Hassan, A. \emph{et al.} Corner states of light in photonic waveguides. \emph{Nat. Photon.} \textbf{13}, 697-700 (2019).	

\bibitem{fraction} Bandres, M. A., Rechtsman, M. C. \& Segev, M. Topological photonic quasicrystals: fractal topological spectrum and protected transport. \emph{Phys. Rev. X} \textbf{6}, 011016 (2016).

\bibitem{defect1} Noh, J. \emph{et al.} Topological protection of photonic mid-gap defect modes. \emph{Nat. Photon.} \textbf{12}, 408-415 (2018).	

\bibitem{topo1} Wang, Y. \emph{et al.} Quantum Topological Boundary States in Quasi-Crystals. \emph{Adv. Mater.} \textbf{31}, 1905624 (2019).

\bibitem{topo2} Wang, Y. \emph{et al.} Topological Protection of Two-photon Quantum Correlation on a Photonic Chip. \emph{Optica} \textbf{6}, 955-960 (2019).

\bibitem{topo3} Wang, Y. \emph{et al.} Topological Protected Quantum Entanglement. \emph{Preprint at http://arXiv.org/abs/1903.03015} (2019).
		
	\end{thebibliography}
\end{document}